# Twitter Reveals: Using Twitter Analytics to Predict Public Protests

Mohsen Bahrami [1,2] *, Yasin Findik [3], Burcin Bozkaya [2] , Selim Balcisoy [3]

[1] *MIT Media Lab, Massachusetts Institute of Technology, Cambridge, MA, USA*

[2] *School of Management, Sabanci University, Istanbul, Turkey*

[3] *Faculty of Engineering and Natural Science, Sabanci University, Istanbul, Turkey*

## Abstract

The right to protest is perceived as one of the primary civil rights. Citizens participate in mass demonstrations to express themselves and exercise their democratic rights. However, because of the large number of participants, protests may lead to violence and destruction, and hence can be costly. Thus it is important to predict such demonstrations in advance to safeguard against such damages. Recent research has shown that about 75 percent of protests that are regarded as legal, are planned in advance. Twitter, the prominent micro-blogging website, has been used as a tool by protestors for planning, organizing, and announcing many of the recent protests worldwide such as those that led to the Arab Spring, Britain riots, and those against Mr. Trump after the presidential election in the U.S. In this paper, we aim to predict protests by means of machine learning algorithms. In particular, we consider the case of protests against the then-president-elect Mr. Trump after the results of the presidential election were announced in November 2016. We first identify the hashtags calling for demonstration from Trending Topics on Twitter, and download the corresponding tweets. We then apply four machine learning algorithms to make predictions. Our findings indicate that Twitter can be used as a powerful tool for predicting future protests with an average prediction accuracy of over 75% (up to 100%). We further validate our model by predicting the protests held in the U.S. airports after President Trump's executive order banning citizens of seven Muslim countries from entering the U.S. An important contribution of our study is the inclusion of event-specific features for prediction purposes which helps to achieve high levels of accuracy.

* Corresponding Author. Tel: +90 216 483 9000
E-mail address: mohsen(at)sabanciuniv.edu

## 1. Introduction

Citizens participate in mass demonstrations to express themselves and exercise their democratic rights. By means of protests, people express their interests, needs, approval or disapproval of a particular situation, and try to bring a better future to their society. Even though a majority of protests have been reported to be peaceful [1], because of the large number of participants in demonstrations, protests may lead to violence and destruction, causing financial damages and/or psychological effects on the society, and hence can be costly in many dimensions.

Thus, it is important to predict such demonstrations in advance to protect against such damages and reduce their expected costs. Accurate and timely prediction of protests help administrative authorities as well as various other stakeholders such as public and private sector organizations, news agencies, financial institutions, etc. to take precautions and/or deal with the consequences of potentially destructive events. Through this kind of analysis, the politicians and social scientists also can delve deeper into the motivating factors, sentiments of the public, socio-economic reasons as to where, when and why masses gather to protest which kinds of events, to understand the nature and progression of protest events. Failing to predict a major protest (false negative) will result in government authorities to respond much later to the unfolding events, possibly causing costs and damages that could have been otherwise avoided. Incorrectly predicting that a protest will happen (false positive) will unnecessarily alert the public and news agencies, hence causing an undue feeling of insecurity.

Research shows that the protests with higher numbers of participants result in a higher probability of succeeding and lower chance of arrest for individuals [2, 3]; consequently, people try to plan and announce the protests in advance to encourage and mobilize greater participation to make the gathering crowds larger. In fact, about 75 percent of protests that are regarded as legal are planned and announced in advance to mobilize a larger group of participants [2].

With approximately two billion users worldwide [4], social media plays an important role in organization of various social events such as gathering for celebrations, parades, riots, protests and similar kinds of collective actions. During the last decade, social media and especially Twitter have been widely used as organization, information and mobilization tools for protests [5-8]. Twitter is arguably the most prominent and popular microblogging and social networking website worldwide. The statistics indicate that the number of monthly active

Twitter users worldwide during the fourth quarter of 2016 averaged 319 million, where this number had been 310 million at the beginning of 2016 [9]. It is worth mentioning that 67 million of the active Twitter users come from the U.S. and Twitter is the most popular microblogging website in the U.S. in 2016 with a penetration rate of 30.6%. This percentage is projected to surpass 35% in 2020 with more than 70 million active users [9]. Twitter users express their points of view and convey information with more than 500 million tweets per day. It is shown that 85% of tweets are related to news [10], and citizen journalists use Twitter to disseminate information and news in real-time, faster than the famous news agencies [11], and even correct if there is misinformation [12].

Previous studies have shown Twitter's key role in recent protests such as those that led to the Arab Spring [3, 11, 13-16], London riots [17-19], Thailand protests [5], Occupy Wall Street [20-22], and Occupy Oakland movements [23]. Other studies in the literature include those that perform tweet content and sentiment analysis during events and protests [5, 10, 24, and 25], investigate Twitter usage at the times of crises and natural disasters [26, 27], describe, model, and interpret the user networks and relationships between social networks as well as social movements [18, 21, 28-31]. Here our focus is on studies that propose models for predicting protests using social media and especially Twitter. We observe two main approaches to prediction: the first based on the properties of online user networks, interactions on social media, and activity cascades [8, 32], and the second approach based on the features driven from aggregated user posts and their contents. In our study, we take the second approach for prediction, for which we also find a significant number of studies in the literature.

Compton et al. [33] perform a content analysis on tweets to find the important ones containing time and place mentions of future protests in order to detect potential protests. Xu et al. [34] use a more extensive method to capture the implicit time and place mentions and they choose Tumblr as their study data source. Muthiah et al. [2] develop a system based on content and linguistic analysis to predict the time interval and place of potential civil unrest. By applying their system to 10 countries in Latin America, they show efforts to detect the time and place of significant protests. Radinsky and Horvitz [35] use a 22-year database and study the sequences of different events to predict whether an event of interest will occur in the future.

Kallus [11] uses the data of more than 300,000 different websites collected by Recorded Future to predict significant protests using machine learning methods. He studies the case of Egypt during the Arab Spring. Steinert-Threlkeld et al. [3] also study the Arab Spring case using about 14 million tweets collected from 16 countries and show that there is a strong statistical relationship between protest activities in a particular day with the level of coordination in its previous day. Korolov et al. [36] study 2015 Baltimore protests. Using tweet content analysis methods, they classify four types of mobilization tweets, namely: sympathy to the cause, awareness of the protest, motivation to take part, and ability to take part [37]. Then they show that there is a correlation between the linear combination of tweet counts of those four types and the actual occurrences of protests.

Ramakrishnan et al. [38] and Doyle et al. [39] both outline the design architecture of the EMBERS system. EMBERS is an automated intelligent system designed to forecast the collective actions such as protests and election outcomes across the countries of Latin America. EMBERS system collects its data using many open data sources such as news agencies and social media especially Twitter. This system uses features driven from message contents and related statistics as well as activity cascade models, then performs features selection by LASSO regression [40] and finally it combines multiple classifiers rather than one prediction model in order to get more reliable results. Korkmaz et al. [41] design a prediction system (incorporated in EMBERS), and using data collected from Twitter and blogs in six Latin American countries, they show that the heterogeneous data sources can collectively increase the accuracy in prediction of future protests.

In this paper, we aim to predict public protests by means of machine learning algorithms using the features extracted from the collected tweets calling for demonstration. In particular, we consider the case of the U.S. post-election protests in 2016, and make predictions for each of the fifty states in the U.S. Furthermore, we test our model by predicting the protests held in the U.S. airports after president Trump's executive order #13769, and attempt to explain that both protest cases were likely motivated by the same reasons.

The series of protests in the U.S. against Donald Trump, known as the anti-Trump protests, started during the 2016 U.S. presidential election campaigns in and outside the U.S., and continued even after Mr. Trump was elected and appointed as president [42]. After the announcement of Mr. Trump's election victory during the early hours of November 9, 2016, a large group of citizens and activists started to call on other people to gather and protest against Mr. Trump's presidency. The main Twitter hashtag [43] created to call citizens to oppose and protest against the president-elect was #NotMyPresident, which was used in millions of tweets. Utilizing the Twitter platform, activists organized and formed mass demonstrations that lasted for about one week, which were held in several major cities of the U.S. including New York, Los Angeles, Boston, Portland, Washington, and Chicago [44].

Another series of protests were held in major airports in the U.S. against President Trump's executive order #13769 of January 27, 2017. According to this executive order, all refugees were blocked from entering the U.S. for 120 days, and the citizens of seven Muslim countries (Iran, Iraq, Libya, Somalia, Sudan, Syria, and Yemen) were banned from entering the U.S. for 90 days. As a result, thousands of people held protests in various airports in the U.S. between January 28-30 [44]. Similarly, Twitter was utilized in this case as an organization and announcement tool, with two main hashtags of #Travelban and #Muslimban.

The main contribution of our work is two-fold: the results of our study confirm that Twitter analytics can help to predict major protests successfully, and we show that the inclusion of regional features driven by reasons and motivators for protests can help to make more accurate predictions.

## 2. Methodology

In order to predict a protest using Twitter data with an online and real-time system, we propose an algorithm that is comprised of five steps. Figure 1 shows the flowchart of the proposed algorithm. Since we collected the data after the protests happened, we already have information without performing two initial steps. As a result, in this study we investigate steps three to five and refer to previous research to explain the methods used for performing the first two steps.

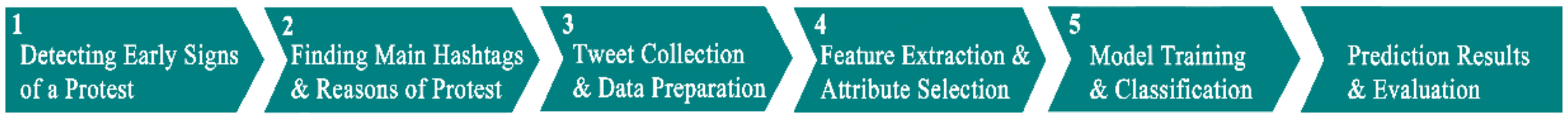

Figure 1. Protest prediction system flowchart

The first step is to search for early signals of a protest in tweet contents. The system should be continuously looking for signals, for example, usage of the words such as "protest", "rally", and similar terms referring to a protest, then consider daily counts of new posts including those keywords [34, 36, 38, and 39]. Whenever the amount of signals in a fixed time interval pass a certain threshold, the system starts the second step.

The second step consists of finding main trending hashtags and reasons of a protest. In the literature, there are different methods to find the reasons of a protest from the collected tweets or user posts [38, 39, and 41]. One approach is to identify the main trending hashtags and then analyze the content of tweets, which use those hashtags. Hashtags identify the topic of tweets [45], help users to easily find tweets with the contents they are interested in, and boost the stream of tweeting. In order to call people to participate in protests, activists usually create one or two main hashtags. Using these hashtags, it is easier for public to express and discuss their sentiment, approval, disapproval, and point of views about that specific protest.

Several hashtags may be created by users during the early hours of tweeting about the protest, but users will coordinate on using a few or even one of them [46]. Using a few hashtags in a large number of tweets is an effective tool for concentrating public attention to a particular topic [46]. There are different approaches to find the main trending hashtags such as the method suggested by Korolov et al. [36] for identifying the relevant hashtags as they emerge.

The first two steps can be performed in a semi-automated way accompanying experts' opinions. For steps 3-5 of the algorithm, we collect and prepare the Twitter data for analysis, then we extract input features / predictors of the model based on our dataset, and finally we select and apply a classifier for prediction.

In what follows, we describe the general approach and then implement our methodology at each step for the specific study case. We make predictions at the state level, which means for fifty U.S. states, we make fifty predictions per day. After extracting the input features and training our classifiers, we output the prediction that whether or not there will be a protest in each state on a particular day. This we represent by an indicator function that has a binary output.

$$\mathbf{1}_i^t = \begin{cases} 1 & if\ there\ will\ be\ a\ protest\ in\ state\ i\ on\ day\ t \\ 0 & if\ there\ will\ be\ no\ protest\ in\ state\ i\ on\ day\ t \end{cases}$$

Finally, we compare the model's prediction output with actual observed protest data [47] to measure the model's performance based on different evaluation measures.

## 2.1 Data Preparation

We aim to investigate the potential power of Twitter in predicting major protests, therefore, a required first step is collecting tweets with contents or signals related to the event of interest. As we described above, hashtags can help to identify the topic of tweets and how they are utilized to boost the stream of twitting. After defining the main hashtag(s), it is possible to collect tweets including them using the Twitter API. We collected the tweets which were tweeted from November 9 to 15, 2016, containing #NotMyPresident and from January 27 to 31, 2017, containing #muslimban and #travelban. Since the prediction is at the U.S. states level, and one needs to know the location of users in order to match tweets with states, only those tweets with known geo-location tags are useful. Twitter users are free to choose their living and tweeting locations to be private or public; thus, as a result of their preferences, the location of a high percentage of tweets is unavailable. We remove such tweets from our data set as we cannot match them with states.

Furthermore, while hashtags hint to the topic of each tweet, they do not guarantee that the collected tweets under a given hashtag all have relevant contents. Invariably a noticeable percentage of the collected tweets have irrelevant contents, and thus they should be removed from the dataset. For example, some users add the top trending hashtags to the end of their advertisement tweets in hopes of increasing the number of users who view their ads. Removing tweets with irrelevant content reduces the data size to about 0.47 million tweets and retweets, which we consider as the population in this study. Figure 2 shows the number of daily tweets (after cleansing) during the post-election protests.

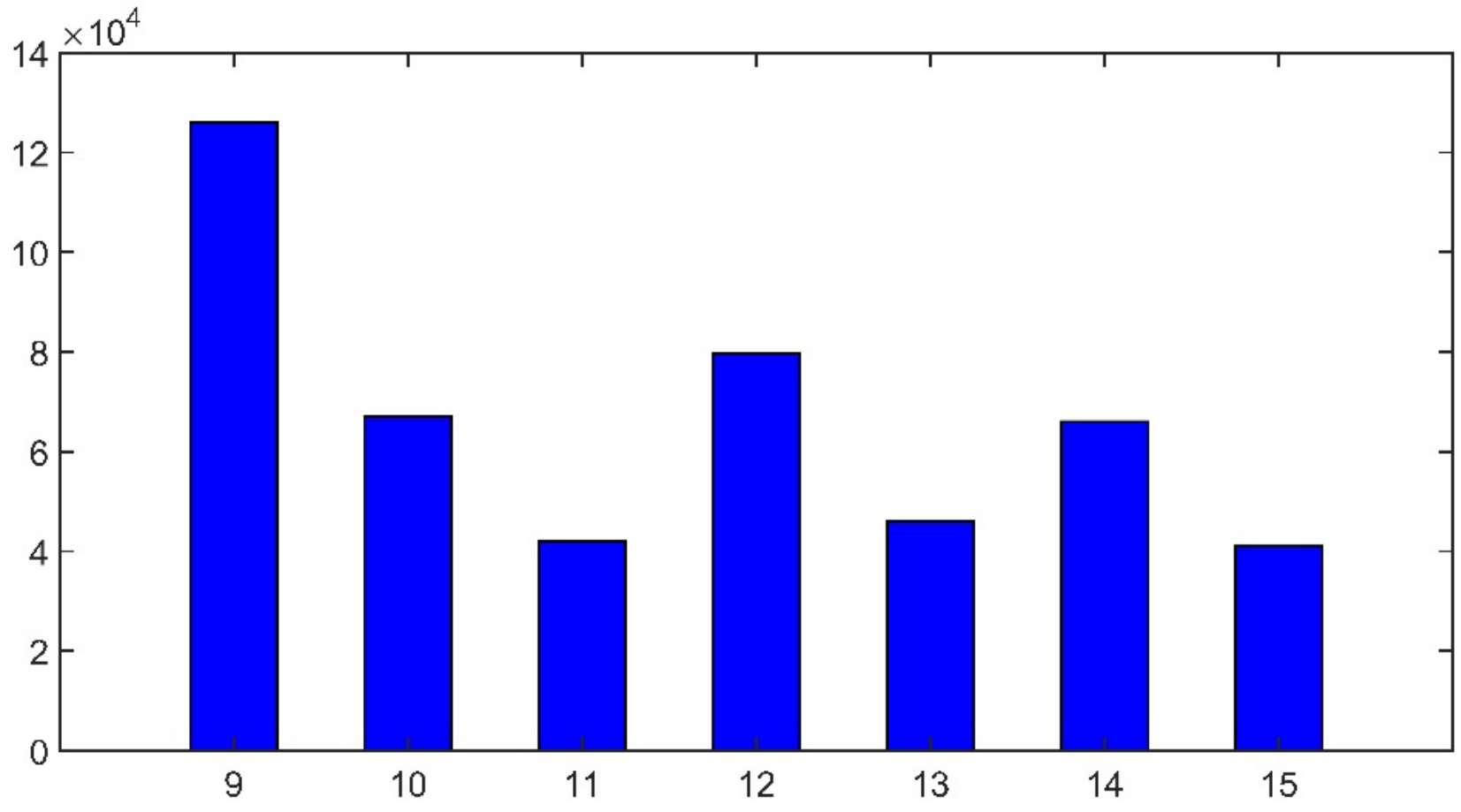

Figure 2. Daily tweet count for the post-election protests after cleansing

Since retweeting can be viewed as an agreement and recommendation mechanism [29, 30], we consider each retweet as a single tweet, independent from the original tweet. For simplicity, we thus refer to “tweets and retweets” only as “tweets” in the sequel.

## 2.2 Feature Extraction

### 2.2.1 Extracting features from the collected tweets

Most of the recent research indicates and agrees on some generic and common tweeting behavior of users before and during most protest and crisis events. These tweeting signals are shown to be common in most of the protest cases studied, independent from where, when, and why they are happening. We use variations of these features as inputs in our prediction model.

The first group of features we extract from the data set pertains to the count of tweets in the context of a protest. As time gets closer to the protest day, an increase in tweeting activity is noticeable.[16] Studies show that the volume of online activities on Twitter is correlated with actual occurrences of offline protests the following day [3, 36]. One significant reason of increase in tweet counts is the users' final attempts to mobilize larger crowds for protest. People are more encouraged/willing to participate in a protest event when they are informed about the large number of potential participants, especially if their family, friends, and neighbors are among those [48, 49]. In our work, we calculate the total tweet count per day, and average tweet count per hour (e.g. average hourly tweeting pace). Tweeting pace can be used instead of tweet count per day when the time intervals of the collected data sets for model training are different.

The second group of features include time, date, and place mentions, which we extracted from the collected tweets using text analysis techniques. These tweets are very important since they explicitly announce the protest times and venues [2, 33, 34, 38, 39, and 41]. There are various ways of mentioning times and dates, which should be considered in the content analysis. Table 1 shows some examples of different ways of time, date, and place mentions. By searching for the mentions in the collected tweets, we derive mention counts for each state (including place) and each particular date (including time).

Table 1. Examples of time, date, and place mentions in tweet contents

| Target word phrase | Example Tweet |
|---|---|
| tomorrow | Protest in my city planned for tomorrow evening #NotMyPresident |
| Indianapolis Saturday | #NotMyPresident Anti-Trump rally planned for Downtown Indianapolis on Saturday |
| November 14th Los Angeles | #LosAngeles high schools will be walking out November 14th 9:15AM. All protests will lead to City Hall. #protest #notmypresident |

Even though these tweets explicitly state the time, date, and place of a potential/planned future protest, it does not mean that the protest will happen for sure or the protest can be predicted only based on these mentions. For example, in the state of Maryland the most mentions calling for protests are on November 15 and 16$^{th}$ where no protests happen whereas there was a protest on November 14th with less protest mentions. The same case also arises for the daily tweet counts.

Afterwards, using the Bag-of-Words method [50], we calculate the average number of violent words [51] per tweet in a daily basis at the state level. Then we perform a sentiment analysis on tweets to find the percentage, and polarity (i.e. strength of the negativity or positivity) of those tweets with negative sentiment, broken down by data and state [52]. The negativity percentage and its polarity are indicators of users' sentiment about the event, which is considered as a predictor of protest, and the average number of violent words per tweet is considered as an indicator of the probability of protest resulting in violence [11].

### 2.2.2. Event Specific Features

There are several reasons that motivate protests and social unrest such as absence of democracy and freedom, political corruption, social injustice, police violence, and unstable economic conditions resulting in a high unemployment rate, poverty, and rising food prices [53, 54]. We argue that tweetting behavioral features are not able to fully represent regional motivators of the protest; moreover Twitter penetration is not geographically comparable for one to safely consider metrics like tweet counts. We propose that the inclusion of the features related to the reasons of each protest would help improve the prediction results. Since protests are special events themselves, each case should be examined specifically. To the best of our knowledge, we are the first to include regional features as motivators for protests used as independent variables in a prediction model.

To construct an improved model, hence we attempt to consider the features that motivate citizens to participate in a particular protest event. In order to find these kinds of features, the reasons and goals of protests should be investigated carefully. Furthermore, we propose that in case of similarity between protests, these features could potentially be used as common predictors for the same kinds of protests.

In the case of post-election protests, which were against Mr. Trump's presidency, we contend that there may be a negative association between the probability of protests happening and Mr. Trump's percentage of votes in each particular state. That is, we argue that those who voted for him are unlikely to participate in a protest against him. Hence, we include Mr. Trump's percentage of votes as a predictor variable in our model.

Figure 3 shows the states where Mr. Trump got the majority of the state votes, as shown in red [55].

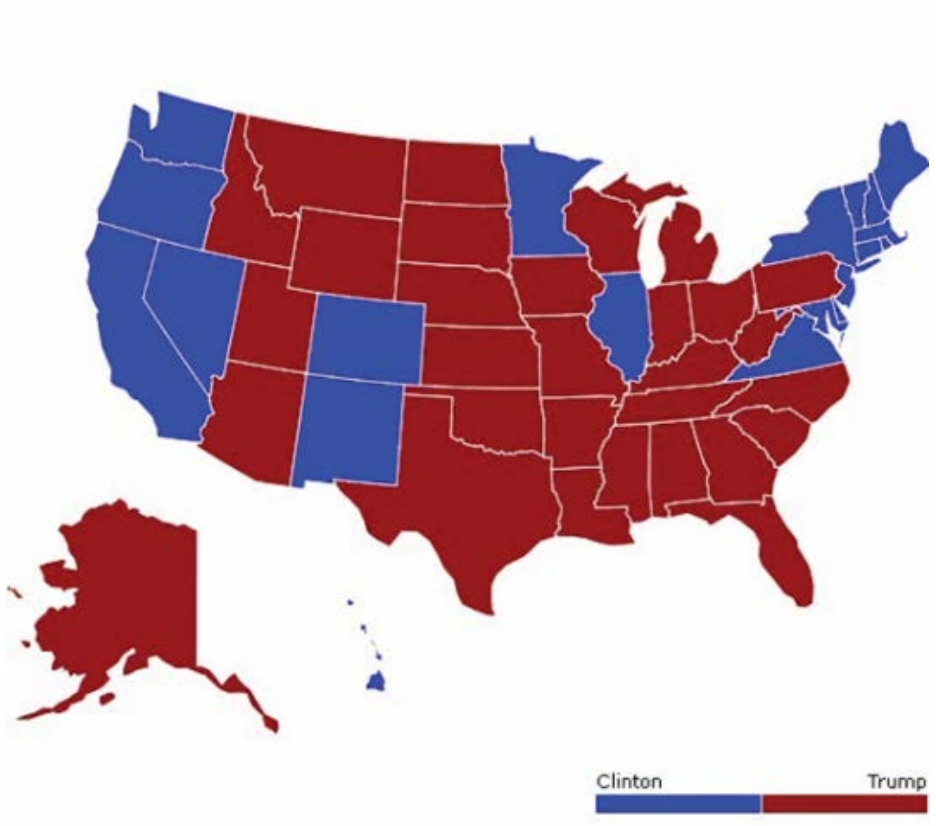

Figure 3. State-level vote distribution map

Total votes in each state is a collection of votes of all counties inside the state; thus, it is important to consider the population effect of large cities and counties as well. To elucidate, in some states such as Texas, even though the majority of the votes is in favor of Mr. Trump, in four counties, Mrs. Clinton has led with a very large margin of vote count compared to her competitors. Investigating the first two days of post-election protests, the corresponding protest maps, and the lead vote map (see Figure 4) [55], we argue that there is also an association between protesting states and the large lead vote sizes in favor of Mrs. Clinton. In order to add large county lead votes to our model, we define a binary variable. The variable is equal to 1 for state *i* if there is a county in that state with leading vote more than a given threshold in favor of Mrs. Clinton, and 0 otherwise. Figure 4 shows the significant lead votes of Mrs. Clinton in each state. The circle size is proportional to the number of votes received by Mrs. Clinton in each county.

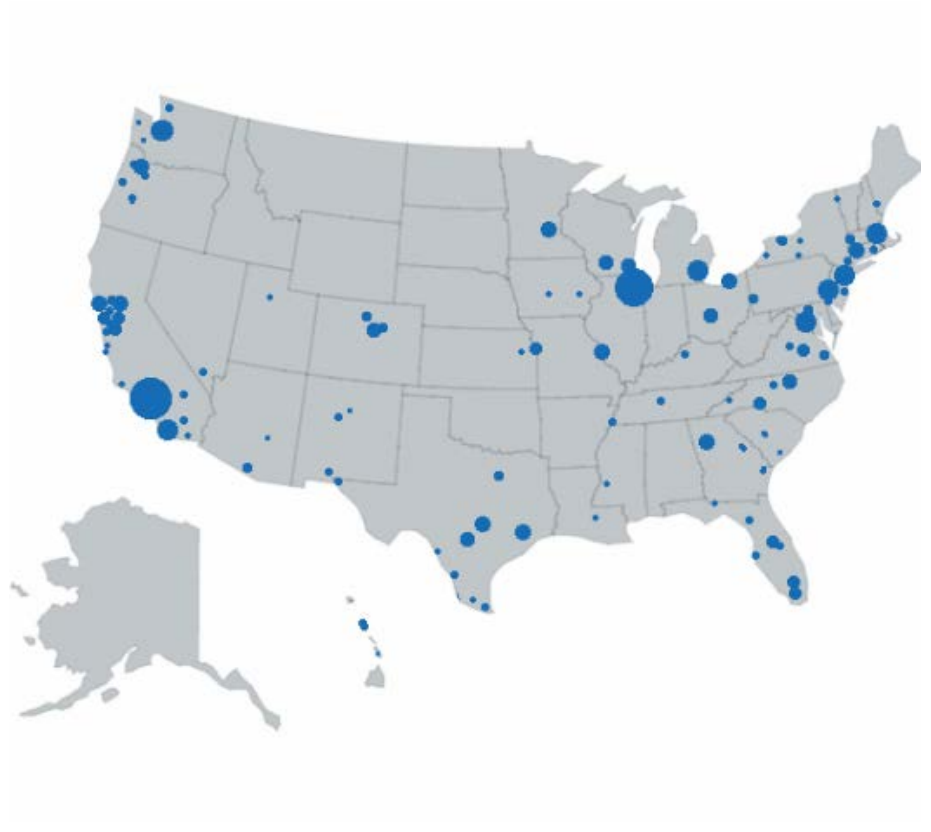

Figure 4. County-based lead vote distribution map

Figures 5 and 6 show the protest maps at the state level in the first two days after the election, on November, 9-10, 2016 [47].

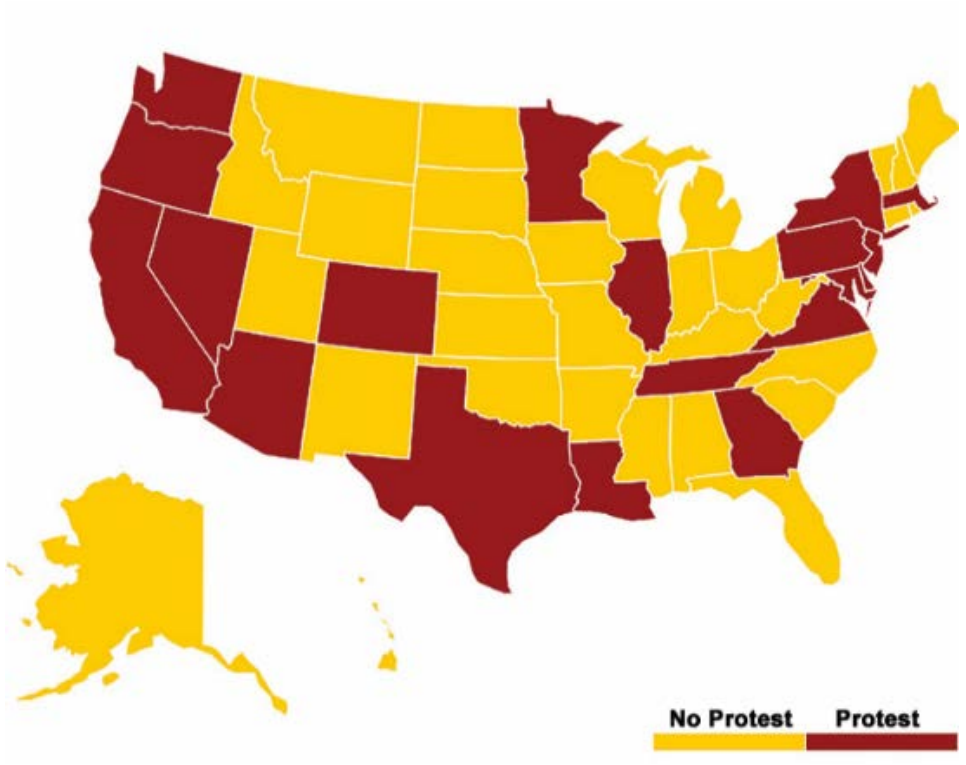

Figure 5. November 9, 2016 post-election protest map

One can visually detect an association between variables (winning vote percentage and the lead vote) visualized in Figures 3 and 4, and the protest maps in Figures 5 and 6.

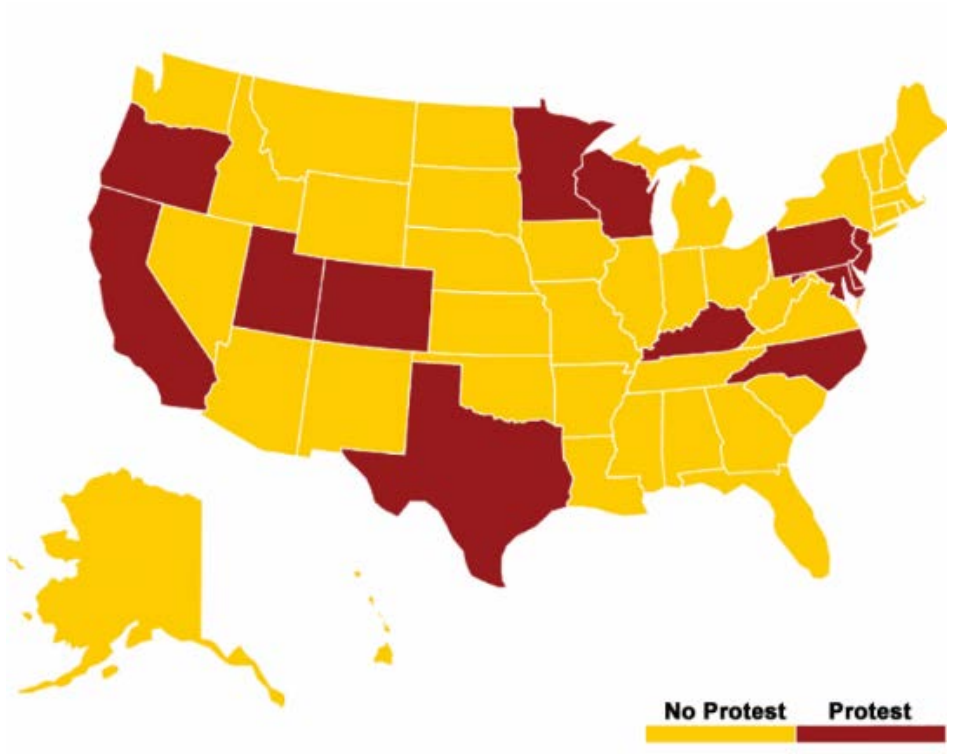

Figure 6. November 10, 2016 post-election protest map

### 2.2.3 Feature Selection

To select the most significant features for our analysis among all extracted ones, before prediction for each next day, we use Wrapper feature selection method [56] with four classifiers namely: C4.5, Naïve Bayes, Logistic Regression (LR), and Support Vector Machines (SVM) [57]. We utilize a bi-directional Best First search [58] on the set of extracted features using the training data with a 10-fold cross validation. We then chose the features which are significant in at least 40% of the folds on average of those four classifiers results. Afterwards, to avoid multi-collinearity, we calculate the pairwise correlation for all features and if there is a high correlation between two features, we remove the one with less significance. For example, the percentage of daily tweets with negative sentiment is highly correlated with total daily number of tweets in each state, as a result we remove the first variable from the set of models' predictors since the Wrapper method results show less significance for the removed feature.

Finally, after the feature selection procedure, we use the following seven variables as the predictors of our model:

- Daily Tweet count (or average tweeting pace if the time intervals are not homogeneous) for each state
- Number of tweets calling for protest in each particular state and the protest date (date, time, and place mentions)
- Average polarity of daily tweets with negative sentiment and average number of violent words per tweet for each state
- Total daily tweet count with negative sentiment in data set as an indicator of overall Twitter users' activity related to the protest topic
- Average number of violent words per tweet per day per state
- Percentage of President Trump's vote in each state
- A binary variable indicating if the county lead vote size is larger than a threshold in each state for Mrs. Clinton

## 2.3 Classifier Selection

As we defined previously, the model's output is a binary variable that indicates whether a protest in each state on a particular day is predicted or not. We have tested various classifiers from different families of classifiers: C4.5, Naïve Bayes, Logistic Regression, and SVM. We obtained the overall best results among these using Logistic Regression with Logit; hence, in the next section we report only the results produced by this classifier. The detailed results with the other three classifiers are provided in the Appendix.

Logistic Regression's output is the probability of protest in each state on a particular date. Although the probability coming from the logistic regression additionally provides a measure of confidence, here we decide to convert the probability value into a binary value. To do this, we need a cutoff value to convert this probability into a binary classification. We consider this property of logistic regression as its strength, since based on the training data one can adjust the cutoff value to get better prediction results.

We allow the algorithm to determine the best cutoff value by utilizing the Single Rule decision tree (OneR method) based on the training data. If the predicted probability is above the cutoff, the model result is converted to 1, meaning that there will be a protest, otherwise it is converted to 0, which is the prediction of no protest.

## 3. Analysis Results

The case we investigate is a chain of protests after the 2016 presidential elections, and unlike some previous studies, the time interval between the announcement of election results and the demonstrations is very short (less than 18 hours). Moreover, there are no tweets recorded prior to the first day of protests; to be able to predict such quickly formed protests, we train our model with each previous day's tweet counts and other predictors mentioned above, for predicting the next day's protests. In our predictions, we handle date/time mentions differently in that we need to consider more days of history simply because the date/time mention might be referring to *any* future date.

Each day represents 50 data points, one for each state; after each day, 50 new points are added to the training data. For example, to predict the $4^{th}$ day of protests, 200 available data points (50 per day before the first protest day until the end of the $3^{rd}$ day) will be used to fit the logistic regression model and then prediction is made for the $4^{th}$ day. Comparing prediction results with the actual protest data, we report the model's performance. Therefore, at the end of the $4^{th}$ day, 250 data points will be available to train the model for $5^{th}$ day's predictions. Our algorithm represents a progressive prediction model, where we fit and tune the model parameters with all available data collected from previous days. (For more details about training data size, refer to Appendix.)

We present the performance results (prediction accuracies) of the two models we trained in Tables 2 and 3. In both tables, we report three predictor performance measures as True Positive Rate (TPR), True Negative Rate (TNR), and Overall accuracy. TPR indicates the success rate in predicting the protests, while TNR indicates the success rate in the prediction of no protests, and the overall accuracy is the weighted average of TPR and TNR based on the number of total protests on a daily basis at the state level. This first model (Model 1 presented in Table 2) excludes the event-specific predictor variables, whereas the second model (Model 2 presented in Table 3) includes all predictor variables.

Table 2. Post-election protest prediction results using features extracted from Twitter (Model 1)

| Predicted date | Nov 11 | Nov 12 | Nov 13 | Nov 14 | Nov 15 | Nov 16 |
|---|---|---|---|---|---|---|
| TPR | 50% | 46.66% | 72.72% | 71.43% | 100% | 100% |
| TNR | 97.5% | 97.14% | 82.05% | 88.37% | 87.5% | 88% |
| Overall accuracy | 88% | 82% | 80% | 86% | 88% | 88% |

Table 3. Post-election protests prediction results using features extracted from Twitter and event specific features (Model 2)

| Predicted date | Nov 11 | Nov 12 | Nov 13 | Nov 14 | Nov 15 | Nov 16 |
|---|---|---|---|---|---|---|
| TPR | 80% | 53.33% | 63.33% | 71.43% | 100% | 100% |
| TNR | 57.5% | 88.57% | 84.61% | 88.37% | 89.58% | 90% |
| Overall accuracy | 62% | 78% | 80% | 86% | 90% | 90% |

We believe that the TPR is more important, first because of the costs associated with predicting a protest as a no-protest, which might be higher than predicting a protest while there will be no protest. Second, protests are rare events, thus if we use a dummy classifier that always predicts no protests during the post-election protests days, the results will be correct in more than 70% of the cases (total accuracy) and 100% correct in predicting of no protests (TNR).

We find that the event-specific variables contribute to the TPR in the early days of the protests, which is quite useful because of the limited number of tweet records during that time frame. As the protests saturate, both models seem to perform similarly for predicting the protests.

We also find that as the size of the training data increases, all three prediction accuracy rates increase simultaneously to a very high success rate. The high success rates produced by these Logistic Regression models confirm that Twitter analytics can be used as a powerful tool to predict the protests.

Another widely used classifier performance evaluation metric is the area under ROC curve [59]. ROC curve indicates the TPR-TFR trade off as the cutoff value changes. Figure 7 shows the ROC curve for models 1 and 2 based on the total prediction results of November 11 to 16. Area under ROC curves (AUC) show that both models are successful in prediction of protests, moreover the results confirm that inclusion of event specific features can increase the performance of the predictive model. (AUC of model 2 is larger.)

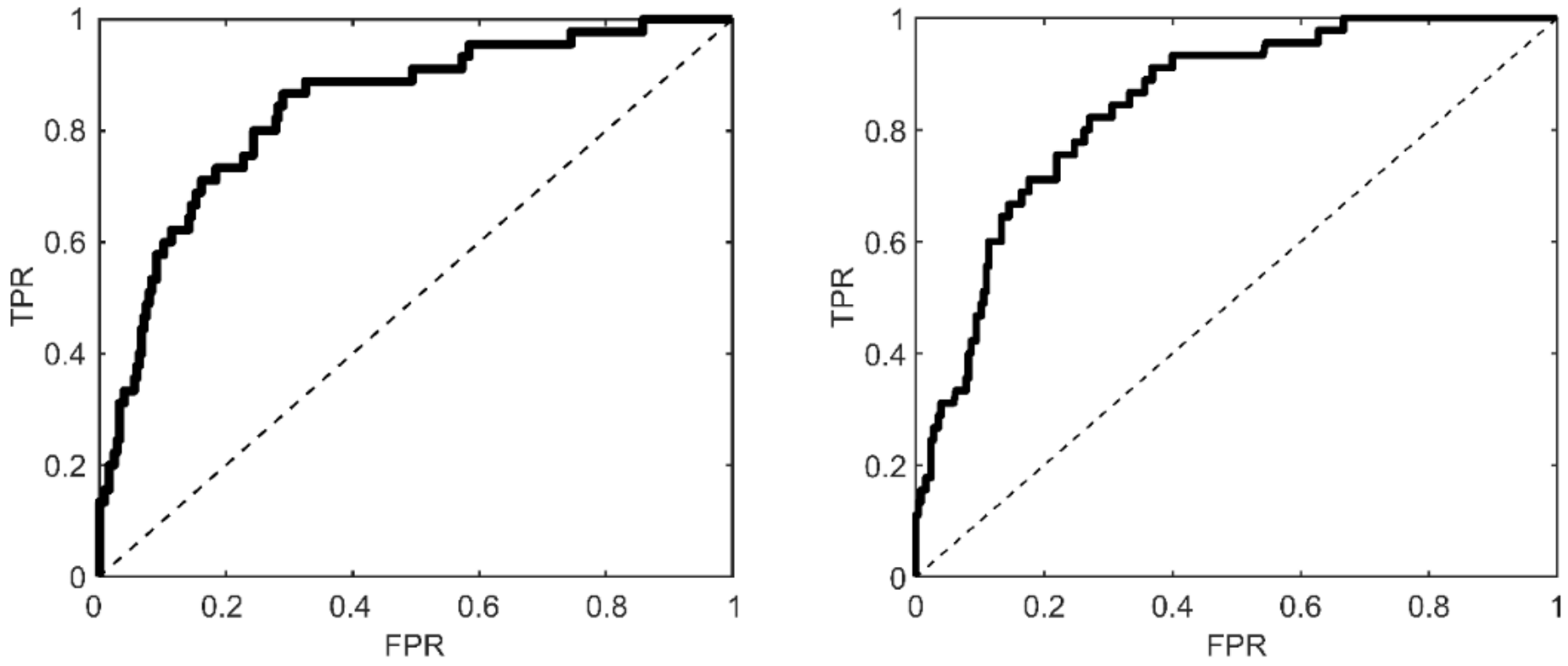

Figure 7. ROC curves of Model 1 (left) and Model 2 (right) based on overall prediction results for post-election protests, AUC1 = 84.07%, AUC2= 84.42%

Next, we use the same two models trained by the post-election protests data of November 10-16, 2016 to predict the protests of January 29, 2017 against the presidential executive order #13769. We were unable to collect enough tweet records for January 29 and 30, and hence we are unable to present prediction results for January 30 and 31.

The models are trained with the post-election protests and used to predict the protests against the executive order #13769. Figure 8 shows the ROC curve for the models 1 and 2 classifiers confirming better performance of model 2 with larger AUC than model 1.

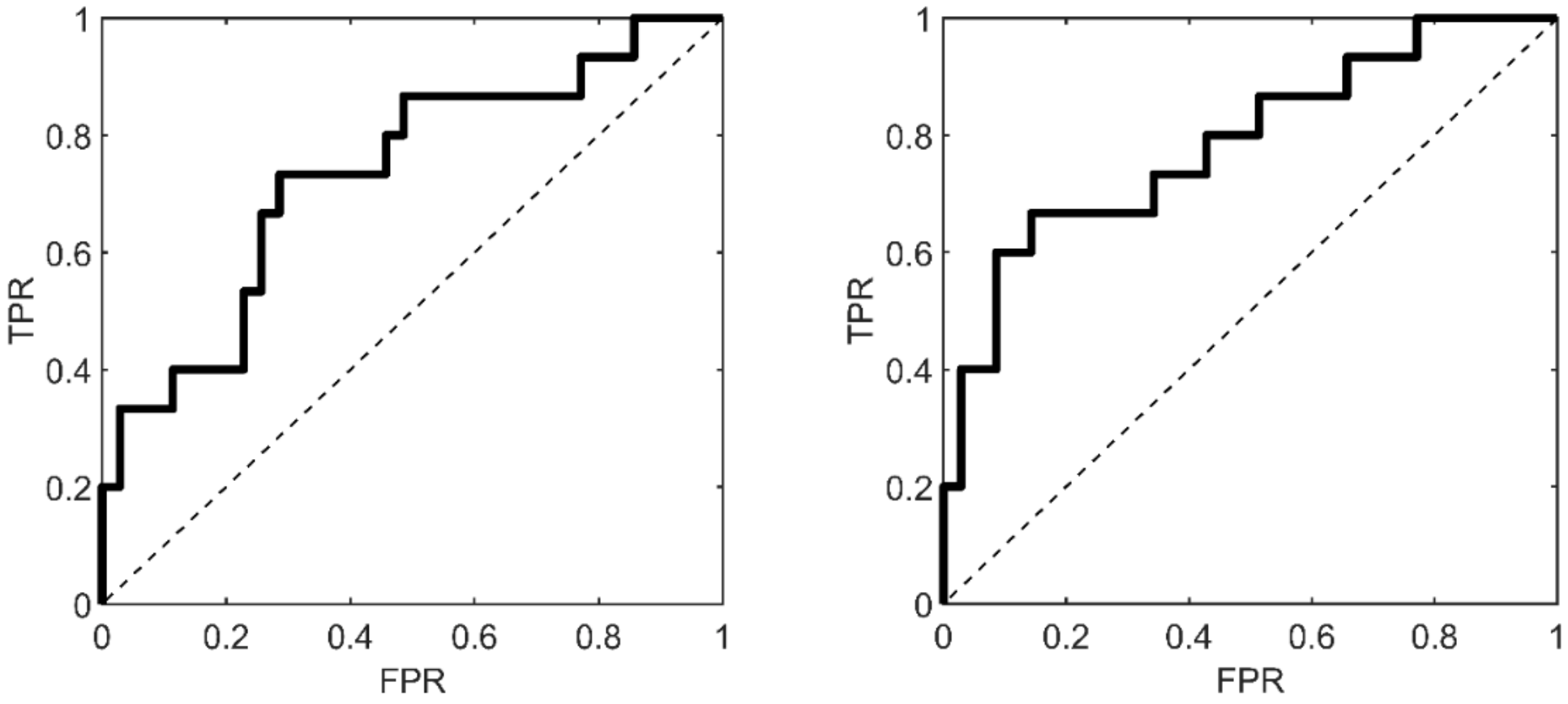

Figure 8. ROC curves of Model 1 (left) and Model 2 (right) used for prediction of protests against the executive order #13769, AUC1 = 73.33%, AUC2 = 78.66%

Based on these results, we find that the post-election and post-travel-ban activity on Twitter along with the most relevant hashtags provide a significant predictive power. Furthermore, we observe that using event specific features further help to increase the predictive power of our model and hence the resulting prediction accuracy values. Cleary, with both approaches, one needs to conduct preliminary steps, that is identifying the most relevant hashtags and also those features that are specific to the event of interest. These may seem to introduce additional effort, but as we discussed before, existing studies can facilitate identification of hashtags and event specific variables.

The results also suggest that there is a similarity between the two cases considered, because the models, trained with voting statistics and post-election tweets, can predict both the post-election protests and the protest against executive order #13769, with reasonably high accuracy. (Results for three baseline methods are provided in Appendix table A4)

Our approach of using the Wrapper method also has further shown that the voting statistics and the post-election specific features always end up being in the set of most effective predictors. Consequently, one may speculate that both cases of protests were motivated by a set of common reasons, yet we believe there might be other features and reasons as well involved in amplifying the chain of protests under study.

Detailed results with different classifiers and more detailed information about features are available in Appendix.

## 4. Conclusion

In this study, we present a classification-based prediction model for predicting mass protests based on Twitter data and tweeting behavior of users. We additionally consider and include event-specific features (e.g. voting statistics) in our model to improve the prediction performance. Both models (with and without event-specific features) predict the post-election protests with high accuracy and true-positive, true-negative rates. In order to show the robustness of our proposed models and the similarity of two cases of mass protests that are potentially related, we have trained our models with one case of protests (i.e. post-election events) and tested them with a second case of protests (i.e. the Muslim-ban events due to presidential executive order #13769). We find satisfying levels of prediction accuracy with the latter case of protests. Our analysis and case study results suggest that the features extracted from Twitter and the models we have developed can potentially be used for protest prediction in the countries where the protests are considered as legal and Twitter is actively used. Furthermore, we note that adding event specific features improves the prediction performance for both cases of protests we studied, which suggests that such features may serve as common motivators or reasons for the related protest events. Finally, our study results emphasize the key role of social media, especially Twitter, in recent protests as an organization and information tool, and that the Twitter has the power of revealing answers for many research questions. For future work, one may further try to validate the proposed models with different events and investigate what other event-specific features can be generated for various events. It may also be possible to employ different machine learning algorithms to improve the prediction accuracy, given the hashtags and tweet content as well as the event-specific features. Finally, it could be worthwhile applying cost-sensitive learning to each event separately, as different costs associated with classification outcomes (false positive, false negative, etc.) may have different impact on algorithm selection as well as prediction accuracies.

## References

1. McLeod, D. M., & Hertog, J. K. Social control, social change and the mass media's role in the regulation of protest groups. Mass media, social control, and social change: A macrosocial perspective, 1999; 305-330.
2. Muthiah, S., Huang, B., Arredondo, J., Mares, D., Getoor, L., Katz, G., & Ramakrishnan, N. Planned Protest Modeling in News and Social Media. In AAAI 2015; pp. 3920-3927.
3. Steinert-Threlkeld, Z. C., Mocanu, D., Vespignani, A., & Fowler, J. (2015). Online social networks and offline protest. EPJ Data Science, 4(1), 19.
4. https://www.statista.com/statistics/272014/global-social-networks-ranked-by-number-of-users
5. Bajpai, K., & Jaiswal, A. A framework for analyzing collective action events on Twitter. In Proceedings of the 8th International ISCRAM Conference 2011.
6. Earl, J., McKee Hurwitz, H., Mejia Mesinas, A., Tolan, M., & Arlotti, A. This protest will be tweeted: Twitter and protest policing during the Pittsburgh G20. Information, Communication & Society, 2013; 16(4), 459-478.
7. Valenzuela, S. (2013). Unpacking the use of social media for protest behavior: The roles of information, opinion expression, and activism. American Behavioral Scientist, 57(7), 920-942.
8. González-Bailón, S., Borge-Holthoefer, J., Rivero, A., & Moreno, Y. (2011). The dynamics of protest recruitment through an online network. Scientific reports, 1, 197.
9. https://www.statista.com/statistics/282087

10. Kwak, H., Lee, C., Park, H., & Moon, S. What is Twitter, a social network or a news media? In Proceedings of the 19th ACM international conference on World Wide Web 2010; pp. 591-600.
11. Kallus, N. Predicting crowd behavior with big public data. In Proceedings of the 23rd ACM International Conference on World Wide Web 2014; pp. 625-630.
12. https://www.theguardian.com/news/datablog/2011/dec/08/twitter-riots-interactive
13. Harlow, S., & Johnson, T. J. The Arab spring| overthrowing the protest paradigm? How the New York Times, global voices and twitter covered the Egyptian revolution. International Journal of Communication, 2011; 5, 16.
14. Bruns, A., Highfield, T., & Burgess, J. (2013). The Arab Spring and social media audiences: English and Arabic Twitter users and their networks. American Behavioral Scientist, 57(7), 871-898.
15. Eltantawy, N., & Wiest, J. B. (2011). The Arab spring: Social media in the Egyptian revolution: reconsidering resource mobilization theory. International Journal of Communication, 5, 18.
16. Lotan, G., Graeff, E., Ananny, M., Gaffney, D., & Pearce, I. (2011). The Arab Spring| the revolutions were tweeted: Information flows during the 2011 Tunisian and Egyptian revolutions. International journal of communication, 5, 31.
17. Cheong, M., Ray, S., & Green, D. interpreting the 2011 London riots from twitter metadata. In Intelligent Systems Design and Applications (ISDA), 12th International Conference on IEEE 2012; pp. 915-920.
18. Gupta, A., Joshi, A., & Kumaraguru, P. Identifying and characterizing user communities on twitter during crisis events. In Proceedings of the 2012 workshop on Data-driven user behavioral modelling and mining from social media 2012; pp. 23-26. ACM.
19. Panagiotopoulos, P., Bigdeli, A. Z., & Sams, S. 5 Days in August–How London local authorities used Twitter during the 2011 riots. In International Conference on Electronic Government. Springer Berlin Heidelberg. 2012; pp. 102-113.
20. Theocharis, Y., Lowe, W., van Deth, J. W., & García-Albacete, G. Using Twitter to mobilize protest action: online mobilization patterns and action repertoires in the Occupy Wall Street, Indignados, and Aganaktismenoi movements. Information, Communication & Society, 2015; 18(2), 202-220.
21. Tremayne, M. Anatomy of protest in the digital era: A network analysis of Twitter and Occupy Wall Street. Social Movement Studies, 2014; 13(1), 110-126.

22. Conover, M. D., Ferrara, E., Menczer, F., & Flammini, A. (2013). The digital evolution of occupy Wall Street. PloS one, 8(5), e64679.
23. Croeser, S., & Highfield, T. Occupy Oakland and# oo: Uses of Twitter within the Occupy movement. First Monday, 2014; 19(3).
24. Bollen, J., Mao, H., & Zeng, X. Twitter mood predicts the stock market. Journal of computational science, 2011; 2(1), 1-8.
25. Hu, Y., Wang, F., & Kambhampati, S. Listening to the Crowd: Automated Analysis of Events via Aggregated Twitter Sentiment. In IJCAI. 2013.
26. Brown, S. Twitter Usage in Times of Crisis. Open Access Journals for School Teachers in Indonesia, 2011; 29.
27. Sakaki, T., Okazaki, M., & Matsuo, Y. Earthquake shakes Twitter users: real-time event detection by social sensors. In Proceedings of the 19th ACM international conference on World Wide Web 2010; pp. 851-860.
28. Mendoza, M., Poblete, B., & Castillo, C. "Twitter under Crisis: Can we trust what we RT?" In Proceedings of the first workshop on social media analytics 2010; pp. 71-79. ACM.
29. Starbird, K., & Palen, L. Pass it on? : Retweeting in mass emergency. International Community on Information Systems for Crisis Response and Management 2010; pp. 1-10.
30. Starbird, K., Muzny, G., & Palen, L. Learning from the crowd: collaborative filtering techniques for identifying on-the-ground Twitterers during mass disruptions. In Proceedings of 9th International Conference on Information Systems for Crisis Response and Management, ISCRAM 2012.
31. Theocharis, Y. The wealth of (occupation) networks? Communication patterns and information distribution in a Twitter protest network. Journal of Information Technology & Politics, 2013; 10(1), 35-56.
32. Cadena, J., Korkmaz, G., Kuhlman, C. J., Marathe, A., Ramakrishnan, N., & Vullikanti, A. (2015). Forecasting social unrest using activity cascades. PloS one, 10(6), e0128879.
33. Compton, R., Lee, C., Lu, T. C., De Silva, L., & Macy, M. Detecting future social unrest in unprocessed twitter data: "emerging phenomena and big data". In Intelligence and Security Informatics (ISI), IEEE International Conference On 2013; pp. 56-60.

34. Xu, J., Lu, T. C., Compton, R., & Allen, D. Civil unrest prediction: A tumblr-based exploration. In International Conference on Social Computing, Behavioral-Cultural Modeling, and Prediction. Springer International Publishing 2014; pp. 403-411.
35. Radinsky, K., & Horvitz, E. Mining the web to predict future events. In Proceedings of the sixth ACM international conference on Web search and data mining 2013; pp. 255-264.
36. Korolov, R., Lu, D., Wang, J., Zhou, G., Bonial, C., Voss, C., ... & Ji, H. (2016, August). On predicting social unrest using social media. In Advances in Social Networks Analysis and Mining (ASONAM), 2016 IEEE/ACM International Conference on (pp. 89-95). IEEE.
37. Van Stekelenburg, J., & Klandermans, B. (2013). The social psychology of protest. Current Sociology, 61(5-6), 886-905.
38. Ramakrishnan, N., Butler, P., Muthiah, S., Self, N., Khandpur, R., Saraf, P., ... & Kuhlman, C. (2014, August). 'Beating the news' with EMBERS: forecasting civil unrest using open source indicators. In Proceedings of the 20th ACM SIGKDD international conference on Knowledge discovery and data mining (pp. 1799-1808). ACM.
39. Doyle, A., Katz, G., Summers, K., Ackermann, C., Zavorin, I., Lim, Z., ... & Lu, C. T. (2014). Forecasting significant societal events using the Embers streaming predictive analytics system. Big Data, 2(4), 185-195.
40. https://en.wikipedia.org/wiki/Lasso_(statistics)
41. Korkmaz, G., Cadena, J., Kuhlman, C. J., Marathe, A., Vullikanti, A., & Ramakrishnan, N. (2015, August). Combining heterogeneous data sources for civil unrest forecasting. In Advances in Social Networks Analysis and Mining (ASONAM), 2015 IEEE/ACM International Conference on (pp. 258-265). IEEE.
42. https://en.wikipedia.org/wiki/United_States_presidential_election,_2016_timeline
43. https://en.wikipedia.org/wiki/Hashtag
44. https://en.wikipedia.org/wiki/Protests_against_Donald_Trump
45. Romero, D. M., Meeder, B., & Kleinberg, J. (2011, March). Differences in the mechanics of information diffusion across topics: idioms, political hashtags, and complex contagion on twitter. In Proceedings of the 20th international conference on World Wide Web (pp. 695-704). ACM.

46. Lehmann, J., Gonçalves, B., Ramasco, J. J., & Cattuto, C. (2012, April). Dynamical classes of collective attention in twitter. In Proceedings of the 21st international conference on World Wide Web (pp. 251-260). ACM.
47. https://www.nytimes.com/interactive/2016/11/12/us/elections/photographs-from-anti-trump-protests.html
48. Granovetter, M. S. (1973). The strength of weak ties. American journal of sociology, 78(6), 1360-1380.
49. Gould, R. V. (1991). Multiple networks and mobilization in the Paris Commune, 1871. American Sociological Review, 716-729.
50. Zhang, Y., Jin, R., & Zhou, Z. H. (2010). Understanding bag-of-words model: a statistical framework. International Journal of Machine Learning and Cybernetics, 1(1-4), 43-52.
51. https://myvocabulary.com/word-list/violence-vocabulary/
52. http://blog.aylien.com/building-a-twitter-sentiment-analysis-process-in/
53. Huihou, A. (2012). The Reasons and Consequences of Political and Social Unrest in Arab Countries. Journal of Middle Eastern and Islamic Studies (in Asia), 6(2), 3.
54. Bellemare, M. F. (2015). Rising food prices, food price volatility, and social unrest. American Journal of Agricultural Economics, 97(1), 1-21.
55. https://www.nytimes.com/elections/results/president
56. Kohavi, R., & John, G. H. (1997). Wrappers for feature subset selection. Artificial intelligence, 97(1-2), 273-324.
57. Friedman, Jerome, Trevor Hastie, and Robert Tibshirani. The elements of statistical learning. Vol. 1. New York: Springer series in statistics, 2001.
58. Dechter, R., & Pearl, J. (1985). Generalized best-first search strategies and the optimality of A. Journal of the ACM (JACM), 32(3), 505-536.
59. https://en.wikipedia.org/wiki/Receiver_operating_characteristic

## Appendix: Detailed Results

Table A 1. Number of data points available for training model for each date predictions

| Date | Nov 11 | Nov 12 | Nov 13 | Nov 14 | Nov 15 | Nov 16 | Jan 29 |
|---|---|---|---|---|---|---|---|
| # of available data points for training | 50 | 100 | 150 | 200 | 250 | 300 | 350 |

Table A 2. Features used as inputs of our logistic regression predictive model after feature selection process ordered by significance given by Wrapper method

| Feature index | Explanation |
|---|---|
| F1 | Time, Date, Place mention count for each day |
| F2 | Count of total daily tweets with negative sentiment |
| F3 | Mrs. Clinton (or President Trump) vote percentage at each state |
| F4 | Average polarity of tweets with negative sentiment per state per day |
| F5 | Average number of violent words per tweets per state per day |
| F6 | Count of daily tweets per state |
| F7 | Existence of lead vote size is larger than threshold in each state for Mrs. Clinton |

Table A 3. Percentage of the overall folds each feature is in the set of best predictors with Wrapper method using four different classifiers

| Feature Index | LR | SVM | Naïve Bayes | C4.5 | Average |
|---|---|---|---|---|---|
| F1 | 90% | 60% | 100% | 30% | 70% |
| F2 | 70% | 50% | 50% | 80% | 62.5% |
| F3 | 70% | 40% | 40% | 70% | 55% |
| F4 | 50% | 60% | 50% | 20% | 45% |
| F5 | 50% | 60% | 50% | 40% | 50% |
| F6 | 40% | 40% | 70% | 50% | 50% |
| F7 | 40% | 20% | 60% | 50% | 42.5% |

Figure A1 shows pairwise Pearson correlation between the features used in our model after feature selection process explained in section 2.2.3

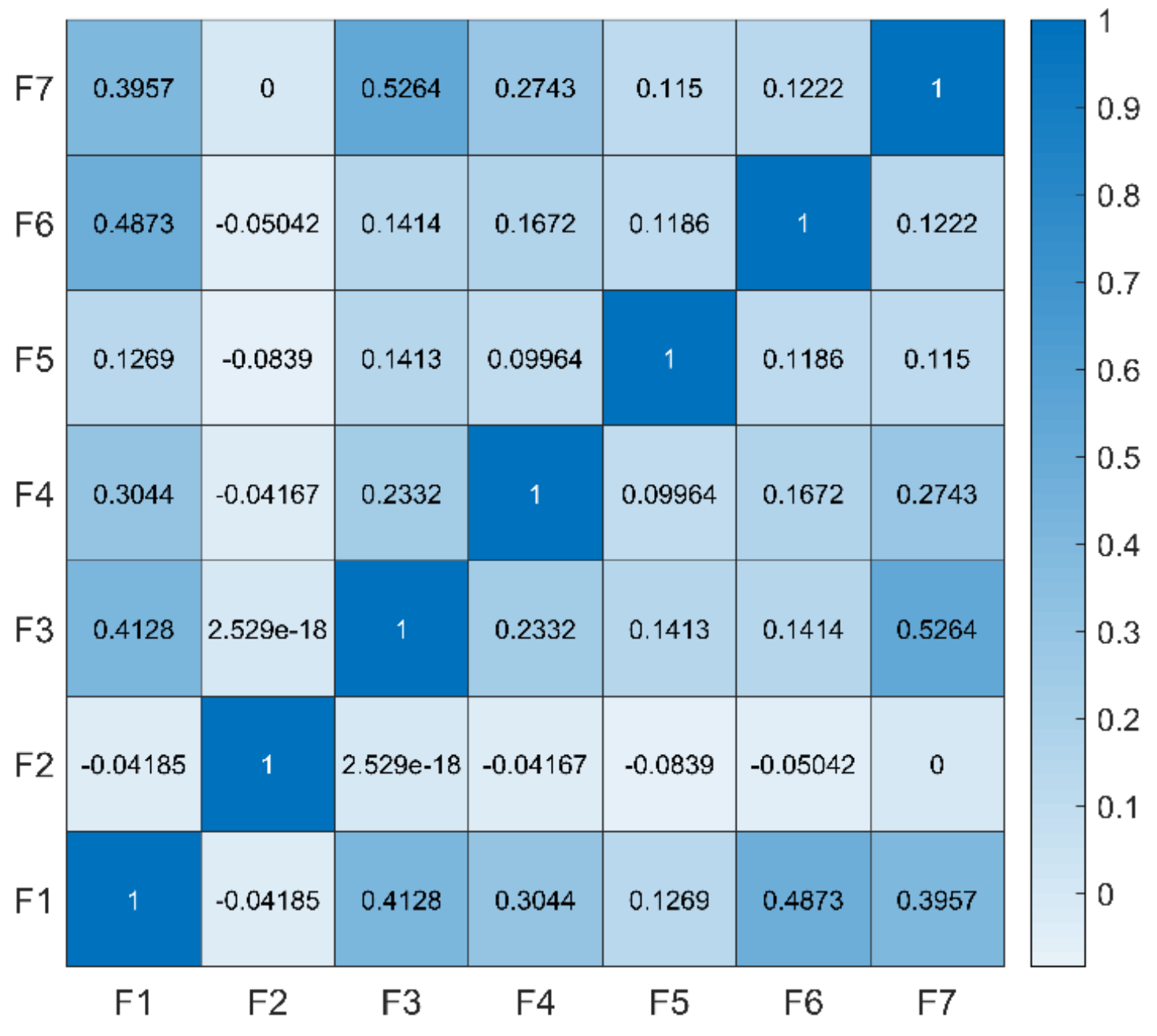

Figure A1. Pairwise Pearson Correlation of predictive features

Table A4 shows how the three most significant features perform in prediction when they are used as single predictors.

Table A 4. Prediction results using each feature (F1, F2, and F3) as single predictor

| Feature | Measure | Nov 11 | Nov 12 | Nov 13 | Nov 14 | Nov 15 | Nov 16 | Jan 29 |
|---|---|---|---|---|---|---|---|---|
| F1 | TPR | 20% | 20% | 27.27% | 42.85% | 50% | 100% | 20% |
| | TNR | 100% | 97.14% | 97.43% | 95.34% | 93.75% | 92% | 100% |
| | ACC | 84% | 74% | 82% | 88% | 92% | 92% | 76% |
| F2 | TPR | 0% | 0% | 0% | 0% | 0% | 100% | 0% |
| | TNR | 100% | 100% | 100% | 100% | 100% | 100% | 100% |
| | ACC | 80% | 70% | 78% | 86% | 96% | 100% | 70% |
| F3 | TPR | 0% | 0% | 36.36% | 42.85% | 50% | 100% | 0% |
| | TNR | 100% | 100% | 97.43% | 93.02% | 89.58% | 100% | 100% |
| | ACC | 80% | 70% | 84% | 86% | 88% | 100% | 70% |

Table A 5. Prediction results with 3 other classifiers using Model 2 features

| Feature | Measure | Nov 11 | Nov 12 | Nov 13 | Nov 14 | Nov 15 | Nov 16 | Jan 29 |
|---|---|---|---|---|---|---|---|---|
| SVM | TPR | 0% | 0% | 63.6% | 42.9% | 50% | 0% | 20% |
| | TNR | 100% | 100% | 87.2% | 97.7% | 97.9% | 96% | 100% |
| | ACC | 80% | 70% | 82% | 90% | 96% | 96% | 76% |
| C4.5 | TPR | 0% | 20% | 63.6% | 71.4% | 100% | 0% | 53.3% |
| | TNR | 100% | 94.3% | 87.2% | 90.7% | 79.2% | 98% | 85.7% |
| | ACC | 80% | 70% | 82% | 88% | 80% | 98% | 76% |
| Naïve Bayes | TPR | 40% | 66.7% | 54.5% | 71.4% | 100% | 0% | 80% |
| | TNR | 87.5% | 85.7% | 89.7% | 90.7% | 89.6% | 90% | 80% |
| | ACC | 78% | 80% | 82% | 88% | 90% | 90% | 80% |